\documentclass[preprint,prc,aps,epsf,nofootinbib]{revtex4}
\usepackage{epsf,times,amsmath,amsfonts,amsthm,graphicx}

\newcommand{\beq}{\begin{equation}}
\newcommand{\eeq}{\end{equation}}
\newcommand{\bea}{\begin{eqnarray}}
\newcommand{\eea}{\end{eqnarray}}

\newcommand{\nn}{\nonumber}
\newcommand{\benn}{\begin{displaymath}}
\newcommand{\eenn}{\end{displaymath}}
\newcommand{\Lu}{L\"{u}scher}
\newcommand{\pr}{~^\mathcal{P}\!\!\!\int} % principal integral
\newcommand{\ig}{\includegraphics}
\newcommand{\tw}{\textwidth}

\def\simge{%  ``greater than about'' symbol
 \mathrel{\rlap{\raise 0.511ex
  \hbox{$>$}}{\lower 0.511ex \hbox{$\sim$}}}}
\def\simle{%  ``less than about'' symbol
 \mathrel{\rlap{\raise 0.511ex
  \hbox{$<$}}{\lower 0.511ex \hbox{$\sim$}}}}
\begin{document}

\preprint{LBNL-62480}
\preprint{DOE/ER/40762-383}

\title{Fitting two nucleons inside a box: exponentially suppressed 
corrections to the \Lu's formula}

\vspace{0.75cm}

\author{Ikuro Sato}
\affiliation{Lawrence Berkeley Laboratory\\
Nuclear Science Division\\
Berkeley, CA 94720 USA}

\author{Paulo F. Bedaque}
\affiliation{University of Maryland\\
Department of Physics \\
College Park, MD 20742 USA}

\date{\today}

\vspace{0.50cm}

% \vspace{10mm}

\begin{abstract}
Scattering observables can be computed in lattice field theory by measuring the volume dependence of energy levels of two particle states. The dominant volume dependence, proportional to inverse powers of the volume, is determined by the phase shifts.   This universal relation (\Lu's formula) between energy levels and phase shifts is distorted by corrections which, in the large volume limit, are exponentially suppressed. They may be sizable, however, for the volumes used in practice and they set a  limit on how small the lattice can be in these studies.
We estimate these corrections, mostly in the case of two nucleons. Qualitatively, we find that the exponentially suppressed corrections are proportional to the {\it square} of the potential (or to terms suppressed in the chiral expansion) and the effect due to pions going ``around the world'' vanishes. Quantitatively, the size of the lattice should be greater than $\approx(5\,\mbox{fm})^3$ in order to keep finite volume corrections to the phase less than $1^\circ$ for realistic pion mass.
\end{abstract}


\maketitle

\section{Introduction}

Recent attempts at studying nuclear interactions using lattice QCD raise an obvious question: what is the minimum size of the lattice that can be used in order to accommodate two nucleons inside it without significant distortion ? This paper aims at answering  this question.

Lattice field theory calculations are performed using imaginary time. This precludes  the calculation of scattering amplitudes in the {\it infinite} volume limit \cite{testa-maiani}. The usual way of obtaining information on scattering amplitudes with lattice techniques is to use the volume dependence of the two-particle energy levels \cite{hamber}. It is intuitively clear that the energy levels of a two-particle system are moved up (down) for a repulsive (attractive) interaction and that this shift, due to the interactions, vanishes in the infinite volume limit. 
In general the volume dependence of the energy levels is a complicated result of the dynamics but if the size of the box is much larger than the range of the interaction between the particles the problem simplifies. 
In this regime we can separate the volume dependences in two categories: the power law dependences, proportional to $1/L^3$ (where $L$ is the size of the box) and exponentially suppressed contributions proportional to $e^{-L/R}$ ($R$ is the range of the interaction). The power law dependence is fully determined by the elastic phase shift \cite{luscher1, luscher2, luscher3} at that energy, through the ``\Lu's formula''. This relation is universal in the sense that it does not depend on the underlying forces between the two particles, only on the phase shift at one particular value of the energy. The exponentially suppressed corrections (ESC) are less universal. For not too small boxes, however, they are dominated by the lightest particle that can be exchanged between the particles; in the case of QCD, those are pions. As those pions are soft, standard effective field theory (chiral perturbation theory) techniques can be used to compute the ESC \footnote{Notice the different level of universality between the power law and the exponential $L$ dependence. The power law dependence is the same for two different underlying theories, as long as they have the same phase shift at that energy level. The ESC will be the same in two different theories  only if they both have the same pattern of chiral symmetry and the same low energy parameters.}. This was indeed recently done, at one loop level, for the case of two pions in the isospin $I=2$ channel \cite{pipi_fv}.

At large enough box volume $L^3$ the power law dependence is dominant and the exponentially suppressed ones can be neglected. In practice however, they may still be sizable and spoil the  lattice extractions of phase shifts using  \Lu's formula. An estimate of these effects in the case of two nucleons is particularly urgent now  as the first unquenched calculations of nucleon-nucleon phase-shifts have appeared \cite{nplqcd-nn}.  In this paper we will discuss how to compute the ESC in general and evaluate them numerically using chiral nuclear effective theory  in order to estimate their sizes in feasible lattice calculations.

\section{\Lu's formula and exponentially suppressed corrections}
\subsection{Infinite volume scattering matrix}

%%%%%%%%%%%%%%%%%%%%%% lippmann %%%%%%%%%%%%%%%%%%%%%
\begin{figure}[t]
\ig[width=0.7\tw]{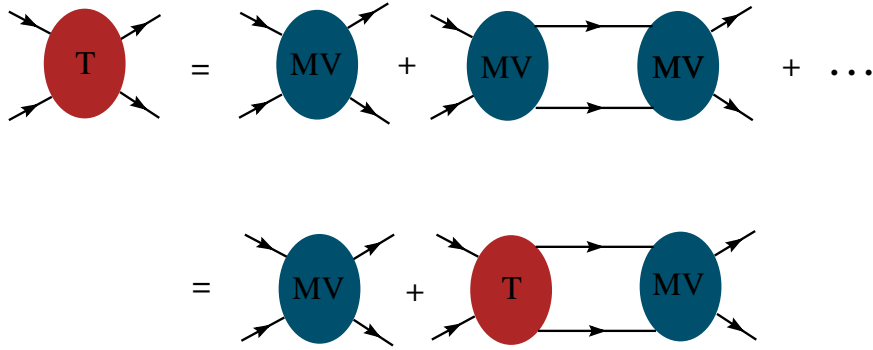}
\caption{Diagrammatic representation of the Lippmann-Schwinger equation.}
\label{fig:LS}
\end{figure}
%%%%%%%%%%%%%%%%%%%%%%%%%%%%%%%%%%%%%%%%%%%%%%%%%%%%%

We first review  the relations
between the potential,  scattering matrix and phase shifts.
The scattering matrix in the non-relativistic case is given by the sum of the (infinite series of)  diagrams shown in the first line of Fig.~\ref{fig:LS}, which  can be summed up by solving the integral equation shown in the second line of Fig.~\ref{fig:LS}. 
This integral equation is simply the Lippmann-Schwinger equation
\begin{equation}
T(\vec{p},\vec{k}) =
-MV(\vec{p}, \vec{k})- M \int \frac{d^3q}{(2\pi)^3}  
\frac{V(\vec{p}, \vec{q})T(\vec{q},\vec{k})}{q^2-k^2-i0},
\end{equation}
where $k=|\vec{k}|$ and $p=|\vec{p}|$ are incoming and outgoing momenta, and $M$ is the rest mass of nucleon.  The on-shell amplitude is given by $T(\vec{k},\vec{k})$\footnote{Our normalization of $T$ includes an extra factor of $M$ compared to the more standard one.}. Projecting onto partial waves we have
\begin{equation}
\label{eq:LS_partial}
T_{lml'm'}(p,k) = -MV_{lml'm'}(p,k) - M \sum_{LM} \int_0^\infty \frac{dq q^2}{2\pi^2} \frac{V_{lmLM}(p,q)T_{LMl'm'}(q,k)}{q^2-k^2-i0}.
\end{equation} 
From now on we'll take the spherically symmetric central potential, i.e., $V_{lml'm'}\sim \delta_{ll'}\delta_{mm'}$. In the infinite volume limit the partial waves uncouple and from now on we will drop the angular momentum indices.

The $i0$ prescription   makes the amplitude complex and it is important only for $q^2\approx ME$, the region of phase space describing the on-shell propagation of the intermediate states over large distances. It is intuitively clear that these are the contributions that will receive the largest finite volume corrections, as we will see below.
It is convenient then to separate from the $T$-matrix the part corresponding to on-shell propagation of intermediate states. For that we rewrite Eq.~\eqref{eq:LS_partial} as
\begin{align}\label{eq:T1}
T(p,k) =&- MV(p,k) - \pr_0^\infty \frac{dq q^2}{2\pi^2} \frac{MV(p,q)T(q,k)}{q^2-k^2}
- i  \underbrace{\pi\int_0^\infty \frac{dq q^2}{2\pi^2} MV(p,q)T(q,k)\delta(q^2-k^2)}_{ MV(p,k)T(k,k)\frac{k}{4\pi}}\nn\\
 =& -MV(p,k) \left( 1+i\frac{k}{4\pi}T(k,k)\right) -  \pr_0^\infty \frac{dq q^2}{2\pi^2} \frac{MV(p,q)T(q,k)}{q^2-k^2},
\end{align} 
where $\pr$ represents a principal value integral.
Since the inhomogeneous term is multiplied by a ($p$ independent) factor, the solution of Eq.~\eqref{eq:T1} will be given by
\begin{equation}
T(p,k) = -K(p,k) \Big[1+i\frac{k}{4\pi}T(k,k)\Big],
\end{equation} 
where $K(p,k)$ satisfies
\begin{equation}
\label{eq:KK}
K(p,k) = MV(p,k) - \pr_0^\infty \frac{dq q^2}{2\pi^2} \frac{MV(p,q)K(q,k)}{q^2-k^2}.
\end{equation} 
The $K$-matrix is real below particle production thresholds since its defining equation is real. Notice that in the definition of $K(p,k)$ the on-shell propagation of intermediate states is not included on the account of the principal value prescription. The on-shell amplitude can now be written as
\begin{equation}
T(k,k) =  \frac{4\pi}{-\frac{4\pi}{K(k,k)} -ik}.
\end{equation} 
Comparing with the standard parameterization of $T(k,k)$ in terms of phase shifts $\delta(k)$
\begin{equation}
T(k,k) = \frac{4\pi}{k}\frac{e^{i2\delta(k)}-1}{2i}= \frac{4\pi}{k\cot\delta(k) -ik},
\end{equation} 
we find 
\begin{equation}
K(k,k) = - \frac{4\pi}{k\cot\delta(k)}.
\end{equation} We can also write the principal value integral as 
\begin{equation}
\pr_0^\infty \frac{dq q^2}{2\pi^2} \frac{V(p,q)K(q,k)}{q^2-k^2}
= \int_0^\infty \frac{dq}{2\pi^2} \frac{q^2V(p,q)K(q,k)-k^2V(p,k)K(k,k)}{q^2-k^2},
\end{equation} 
which is sometimes more convenient, especially in numerical calculations.

\subsection{Finite volume case}

There are a few changes in the derivation of the Lippmann-Schwinger equation as we go from infinite space to a finite box\footnote{Of course, there is no real scattering and/or asymptotic states at finite volume, but we will still call the finite volume version of $T$ the finite volume scattering matrix. }. First,  the allowed momenta of intermediate particles are restricted to discrete values $\vec{q} = 2\pi \vec{n}/L$, $\vec{n} \in \mathbb{Z}^3$.
Second, the projection onto partial waves is more complicated  as the shape of the box breaks rotational symmetry and higher partial waves mix even when the potential is spin independent and spherically symmetric.
The mixing is however suppressed by two effects. The first is that the higher partial waves play little role at low energy scatterings, i.e., $\delta_l(k)\sim k^{2l+1}$ at small $k$. In the case of s-wave, for instance, the largest contamination comes from the $l=4$ partial wave, whose phase shifts are no greater than $2^\circ$ for $0<k<300$\,MeV. The second effect suppressing the partial wave mixing is the approximate orthogonality between the spherical harmonics at finite but large volumes.
For these reasons, the partial wave mixing will be disregarded from now on. Finally, the masses and potentials are also changed from their infinite volume values. Those changes are exponentially suppressed and, as we will argue below, are also suppressed in the chiral expansion, a point that will be further discussed below. In this section, we will keep the infinite volume values of $M$ and $V$. 

We arrive then at the equation defining ${\bf T}$, the {\it finite} volume analogue of the $T$-matrix,
\begin{equation}
\label{eq:T_def}
{\bf T}(p,k) = -MV(p,k) -\frac{1}{L^3}\sum_{\vec{q} = 2\pi\vec{n}/L} \frac{MV(p,q){\bf T}(q,k)}{q^2-k^2}.
\end{equation} 
Assuming that the value of $k$ does not coincide with any of the allowed values of $q$ there is no need for the $i0$ prescription, which makes the finite volume scattering matrix real.
We now use a fundamental result of finite volume momentum sums \cite{sharpe_moving}:  if the summand has no singularities in the real axis and decays to zero fast enough at infinity, the difference between the sum and the integral is exponentially small at large $L$. This result follows from the Poisson summation formula
\begin{align}
\label{eq:poisson}
\frac{1}{L^3}\sum_{\vec{q}= 2\pi\vec{n}/L} f(\vec{q}) =& 
\int \frac{d^3q}{(2\pi)^3} f(\vec{q})
+\sum_{\vec{n}\neq 0, \vec{n}\in\mathbb{Z}^3}\int\frac{d^2q}{(2\pi)^3}f(\vec{q}) e^{i L \vec{q}\cdot\vec{n}},
\nn\\
=& \int \frac{d^3q}{(2\pi)^3} f(\vec{q}) + \mathcal{O}(e^{-mL}),
\end{align} 
where $m$ is the characteristic scale of $f$.  The summand in Eq.~\eqref{eq:T_def} has however a singularity at $q=k$ and the Poisson summation formula cannot be directly used.  We  isolate this singularity by performing similar steps taken above with the infinite volume scattering matrix. We separate the singular term corresponding to on-shell propagation by writing 
\begin{align}
{\bf T}(p,k) = -MV(p,k) &-\frac{1}{L^3}\sum_{\vec{q} = 2\pi\vec{n}/L} \frac{MV(p,q){\bf T}(q,k)-MV(p,k){\bf T}(k,k)}{q^2-k^2}\nn\\
&  -M V(p,k){\bf T}(k,k) \frac{1}{L^3}\sum_{\vec{q} = 2\pi\vec{n}/L} \frac{1}{q^2-k^2}.
\label{bf_T}
\end{align}
The first summand is now regular and the sum can be replaced by integrals
\begin{align}
&\frac{M}{L^3}\sum_{\vec{q}=2\pi\vec{n}/L} \frac{V(p,q){\bf T}(q,k)-V(p,k){\bf T}(k,k)}{q^2-k^2} \nn\\
&=
M\pr \frac{d^3q}{(2\pi)^3} \frac{V(p,q){\bf T}(q,k)}{q^2-k^2} -
M\pr \frac{d^3q}{(2\pi)^3} \frac{V(p,k){\bf T}(k,k)}{q^2-k^2} 
+ {\bf F}(p,k),
\end{align}
with the difference between the sum and the integral  ${\bf F}(p,k)$ being an exponentially small quantity:
\begin{equation}
{\bf F}(p,k)\equiv M\Big( \frac{1}{L^3} \sum_{\vec{q}}
-\int \frac{d^3q}{(2\pi)^3} \Big) 
\frac{V(p,q){\bf K}(q,k)-V(p,k){\bf K}(k,k)}{k^2-q^2}.
\label{bf_F}
\end{equation}
The integrand in ${\bf F}$ is regular, thus
the Poisson summation formula can be applied to Eq.~\eqref{bf_F} to yield
\begin{equation}
{\bf F}(p,k) =-M\sum_{\vec{n}\neq 0,\vec{n}\in\mathbb{Z}^3} 
\int_0^\infty \frac{dq q}{2\pi^2} \frac{\sin(|\vec{n}| q L)}{|\vec{n}|L}
\frac{V(p,q){\bf T}(q,k)-V(p,k){\bf T}(k,k)}{q^2-k^2}.
\end{equation}
The second sum in Eq.~(\ref{bf_T}) includes the power law corrections. The sum itself is a universal function of $k$, independent of the potential. We give it a name by defining
\begin{equation}
\frac{1}{4\pi^2 L}{\bf S}\Big(\frac{k^2L^2}{4\pi^2}\Big)\equiv 
\Big(\frac{1}{L^3}\sum_{\vec{q} = 2\pi\vec{n}/L} -\int \frac{d^3q}{(2\pi)^3} 
\Big)\frac{1}{q^2-k^2}. 
\end{equation} 
We can now write Eq.~(\ref{bf_T}) as 
\begin{equation}\label{eq:T_2}
{\bf T}(p,k) = -MV(p,k) \Big[ 1+  {\bf T}(k,k)\frac{1}{4\pi^2 L}{\bf S}\Big(\frac{k^2L^2}{4\pi^2}\Big) \Big]
-  \pr \frac{d^3q}{(2\pi)^3} 
\frac{MV(p,q){\bf T}(q,k)}{q^2-k^2} + {\bf F}(p,k).
\end{equation} 
The solution of Eq.~(\ref{eq:T_2}) is given by
\begin{equation}
\label{eq:Kvpk}
{\bf T}(p,k) = -\Big[ 1+ {\bf T}(k,k)\frac{1}{4\pi^2 L}{\bf S}\Big(\frac{k^2L^2}{4\pi^2}\Big)\Big] \mathbb{K}(p,k),
\end{equation} 
where $\mathbb{K}(p,k)$ and $\mathbb{F}(p,k)$ are, in their turn, defined by
\begin{equation}
\mathbb{K}(p,k) = MV(p,k) - \pr \frac{d^3q}{(2\pi)^3} 
\frac{MV(p,q)\mathbb{K}(q,k)}{q^2-k^2} + \mathbb{F}(p,k),
\label{mathbb_K}
\end{equation} 
and
\begin{equation}
\mathbb{F}(p,k)
=-M\sum_{\vec{n}\neq 0,\vec{n}\in\mathbb{Z}^3} 
\int_0^\infty \frac{dq q}{2\pi^2} \frac{\sin(|\vec{n}| q L)}{|\vec{n}|L}
\frac{V(p,q)\mathbb{K}(q,k)-V(p,k)\mathbb{K}(k,k)}{q^2-k^2}.
\label{mathbb_F}
\end{equation} 
$\mathbb{K}(p,k)$ is the a finite volume generalization of the infinite volume $K(p,k)$ that includes the ESC, but not the power law finite volume corrections. In fact, dropping the exponentially suppressed term $\mathbb{F}(p,k)$ from Eq.~\ref{mathbb_K}, we recover the defining equation for $K(p,k)$ (Eq.~\ref{eq:KK}). The poles of the ``on-shell'' amplitude ${\bf T}(k,k)$
\begin{equation}
\label{eq:Kvkk}
{\bf T}(k,k) = -\frac{4\pi }{\frac{4\pi}{\mathbb{K}(k,k)}
+\frac{1}{\pi L}{\bf S}(\frac{k^2L^2}{4\pi^2}) }.
\end{equation} 
determine the energy levels in the finite box.

In the large $L$ limit, the finite volume correction $\mathbb{F}$ 
goes to zero (exponentially fast) and 
$\mathbb{K}$ is equal to the infinite volume $K$-matrix defined in 
Eq.~\eqref{eq:KK}.
Then the relation between the energy level shifts and scattering parameters,
the \Lu's formula, is recovered:
 \beq\label{eq:luscher}
 -\frac{4\pi}{K(k,k)} = k \cot\delta(k) =
\frac{1}{\pi L}{\bf S}(\frac{k^2L^2}{4\pi^2}).
 \eeq 
This formula includes the power law correction to the energy levels in a box, but not the ESC. This can be seen more explicitly by approximating, for small $k$, $k\cot\delta(k)$ by $-1/a_0$ (where $a_0$ is the scattering length) and $\mathcal{S}(\frac{k^2L^2}{4\pi^2}) \approx -1/(L^3 k^2)$ to find
\beq
 E = \frac{k^2}{M}=\frac{4\pi a_0}{ML^3} \left[ 1+\mathcal{O}(\frac{a_0}{L}) \right].
\eeq
 
If we include $\mathbb{F}$, the difference
between the infinite and finite volume matrices, $\Delta K = \mathbb{K}-K$,
provides corrections to this relation as follows
\begin{equation}
k\cot\delta(k) \Big[1-\frac{\Delta K(k,k)}{K(k,k)} 
+O\Big(\frac{\Delta K(k,k)^2}{K(k,k)^2}\Big)\Big] = \frac{1}{\pi L}
{\bf S}\Big(\frac{k^2L^2}{4\pi^2}\Big)
\label{Luscher_ESC}
\end{equation}
and $\Delta K$ satisfies the  equation,
\begin{equation}
\Delta K (p,k)= \mathbb{F}(p,k) - M \pr_0^\infty \frac{d^3q}{(2\pi)^3}
\frac{V(p,q)\Delta K(q,k)}{q^2-k^2}.
\label{DK}
\end{equation}
Equation~\eqref{Luscher_ESC} is one form of the \Lu's formula with ESC associated with the scattering matrix due to the finite volume.
Further approximations are possible if the pole is close to free particle levels, i.e., $k^2\approx (2\pi n/L)^2$.
 
Let us recapitulate we have done so far. The large (power law) difference between the finite and infinite volume scattering amplitude comes from the singular $q^2=k^2$ region of the integral. Physically, it corresponds to the kinematics where both intermediate particles are on-shell and can propagate far away, explore the lattice and ``notice'' it is finite. The $\mathbb{K}$-matrix is defined by subtracting this term, so the difference from the infinite volume $K$-matrix is suppressed by terms of order $e^{-m L}$.
The procedure we followed was to isolate the singular term in the finite volume ${\bf T}$ matrix, relate the rest to the $\mathbb{K}$-matrix and the phase shifts and treat separately the singular region. The contribution from the singular region, where the power law $L$ dependence resides, does not depend on the particular interaction, it is an universal function describing the phase space for the two intermediate particles to be on-shell simultaneously and can be computed (numerically in general, but analytic formulae are available in limiting cases). This way a relation between the phase shifts and the energy levels in a box are found.

Besides the ESC appearing in the scattering matrix, there are  ESC to the potential and particle masses themselves. They will be discussed in the next section. 
\subsection{No order $V$ (pions ``around the world'') effect}

When the nuclear potential $V$ is weak, Eqs.~\eqref{mathbb_K} and~\eqref{mathbb_F} show that
$\Delta\mathbb{F}(k,k)$ is quadratic in the potential. It might seem surprising that there is no ESC linear in $V$ coming from pions being exchanged after ``wrapping around the world''. This can be understood in simple terms. 

In momentum space, the potential generated  at finite volume by, for instance, one pion exchange, is the same as in the infinite volume limit. In fact, it is essentially given by the solution of the Klein-Gordon equation in momentum space. The potentials in position space however depend on the volume.
In infinite space we have
\begin{equation}
V(\vec{r}) = \int\frac{d^3q}{(2\pi)^3} e^{i \vec{q}.\vec{r}} V(\vec{q}).
\end{equation} 
but at finite volume we have instead
\begin{align}\label{eq:V_finite}
{\bf V}(\vec{r}) =& \frac{1}{L^3}\sum_{\vec{q}=2\pi\vec{n}/L}
e^{i \vec{q}\cdot\vec{r}} V(\vec{q})\nn\\
=& \sum_{\vec{n}\in \mathbb{Z}^3} \int\frac{d^3q}{(2\pi)^3}
e^{i \vec{q}\cdot(\vec{r}+\vec{n}L)} V(\vec{q})\nn\\
=& \sum_{\vec{n}\in \mathbb{Z}^3} V(\vec{r}+\vec{n}L).
\end{align} 
The first ($\vec{n}=0$) term in Eq.~\eqref{eq:V_finite} reproduces the infinite volume result, the remaining ones are finite volume exponentially suppressed corrections to it. We see then that there are indeed contributions to the potential between two particles in a periodic box coming from ``pions wrapping around the world''. Let us now compute, in first order perturbation theory, the energy shift due to the interaction. For simplicity, let us assume the unperturbed states correspond to plane waves with zero momentum. The energy shift is given by
\begin{align}
\Delta E =& \int_{L^3} d\vec{r}_1 d\vec{r}_2 \frac{1}{L^{3}} \frac{1}{L^{3}} {\bf V}(\vec{r}_1-\vec{r}_2)\nn\\
=& \frac{1}{L^{3}}\int_{L^3} d\vec{r} ~ {\bf V}(\vec{r})\nn\\
=& \frac{1}{L^{3}}\sum_{\vec{n}\in \mathbb{Z}^3} \int_{L^3} d\vec{r}  ~V(\vec{r}+\vec{n}L)\nn\\
=& \frac{1}{L^{3}}\int d\vec{r}~ V(\vec{r})\nn\\
=& \frac{4\pi a_0}{L^{3}},
\end{align} 
where $a_0$ is the scattering length (at infinite volume). The last line above shows the leading power law dependence with the volume contained in  the \Lu's formula. We see then that the ESC to the potential compensate for the fact that the integrations over space are limited to the finite box and the final result, when expressed in terms of the infinite volume quantity $a_0$, contains no ESC.

\section{ $^1S_0$ nucleon-nucleon scattering}

In this section we will estimate numerically the size of the ESC in the case of spin-singlet nucleon-nucleon interactions. It is important to estimate them not only in the case of realistic pion masses but also for the current lattice calculations with pion masses in the $m_\pi=300-500$ MeV range. This can only be done if the behavior of nuclear forces at large quark masses is known. In principle, the nuclear chiral effective theory provides this extrapolation \cite{martin_extra,ulf_extra1,ulf_extra_2}. In practice, these extrapolations are somewhat hindered by the poor knowledge  of some low energy constants and the slow convergence of the chiral expansion so the results we present for non-realistic pion masses should be taken with a grain of salt. Further lattice QCD/effective theory work in the near future should improve the situation markedly.

In the chiral expansion, the different contributions to  the potential  can be divided in short range (with range on QCD scales $r\sim 1/\Lambda_{\rm QCD}$) and long distance (with range on the pion  scale $r\sim 1/m_\pi$)\cite{bedaque_bira_review,bedaque_all_review,phillips_review}. They generate ESC suppressed by factors of, respectively, $e^{-\Lambda_{\rm QCD} L}$ and $e^{-m_\pi L}$. Clearly, the largest one comes from the long distance potential. At leading order in the chiral expansion, the long distance potential is given by one-pion exchange which does not receive finite volume corrections since it is defined by tree diagrams. At higher orders in the chiral expansion there are contributions coming from two-pion exchange. They decay at large distances as $e^{-2m_\pi r}$ and, consequently, generate finite volume corrections proportional to $e^{-2m_\pi L}$. The two-pion exchange contribution to the energy shifts is thus  suppressed by both a factor of $e^{-2m_\pi L}$ and a factor of $m_\pi^2/4\pi f_\pi^2$ coming from the chiral expansion. Also, at the next order of the chiral expansion there are vertex corrections that renormalize the strength of the pion-nucleon coupling and that receive (exponentially small) finite volume corrections. We will not include any of these higher orders  effects to the nuclear forces in our estimates as are they are suppressed in the chiral expansion.
% (or by higher powers of $e^{-m_\pi L}$).

 The short distance part of the potential is described, in the effective theory approach, by contact terms. The contact terms containing no derivatives form a power series on the quark masses or, equivalently, on $m_\pi^2$:
\beq\label{eq:contact}
\mathcal{L}_{NN}^{\rm no\ der.} = (C_0^0 + m_\pi^2 C_0^2+\cdots)(N^\dagger \tau_2 N^*) (N^T \tau_2N),
\eeq where $N$ is the nucleon field and the matrices ${\bf \tau}$ act on isospin space. 
The natural sizes for  the contact terms are 
\bea\label{eq:nda}
C_0^0 &\sim& \frac{4\pi}{M\Lambda},\nn\\
C_0^2 &\sim& \frac{4\pi}{M\Lambda^3}
\eea 
where $\Lambda$ is the high momentum scale where the effective theory breaks down, $\Lambda \approx 500$ MeV. Unfortunately, from the experimental values of the phase shift we have access only to the combination $C_0=C_0^0 + m_\pi^2 C_0^2$, not to $C_0^0$ and $ C_0^2$ individually \footnote{Processes involving emission/absorption of a pion do distinguish these operators.}. As such we can only  make reasonable assumptions based on  naive dimensional analysis (Eq.~(\ref{eq:nda})) about the values of $C_0^0$ and $ C_0^2$ separately, consistent with the  value  of $C_0$ at the physical pion mass determined by fitting experimental phase shifts. 

The next-to-leading corrections to the short distance potential come from the term
\beq
\mathcal{L}_{NN}^{\rm \ two\ der.} = C_2 (N^\dagger \tau_2 \nabla^2 N^*) (N^T \tau_2N)+h.c.
\eeq  
The long range part of the potential also receives corrections contributing at the same order as $C_2$. Both $C_0$ and $C_2$ contact terms are fit to reproduce physical phase shifts of nucleon-nucleon scattering as is discussed in following subsections.
%In practice they are numerically smaller and we will disregard them in our estimates. Our calculation can then be seen as a ``half-next-to-leading'' order calculation. 

%It is well known that the s-wave nucleon-nucleon interactions are fine tuned in the real world and that the scattering lengths, specially in the spin singlet case we are considering, are unnaturally large. That means that the potential is strong, almost strong enough to generate a bound state. This fine tuning is easily destroyed by any change of the theory as, for instance, a change in the quark masses. Without the fine tuning the nuclear potential is actually quite weak. 
%
%When lattice spatial size $L$ is about $1/m_\pi$ or smaller, nucleons at 
%position $\vec{x}$ interact with nucleons at $\vec{x}+\hat{l}L$ via contact 
%terms due to periodic boundary conditions in the space.
%Needless to say, this is an unwanted situation because these fictious 
%interactions can vary energy level significantly.
%On the other hand, when $m_\pi L$ is larger than $\sim 5$,
%interaction with particles in the ``nearlest box'' is dominated by 
%one-pion-exchange, in which case the
%energy distortion is under controll and analytical formula for ESC is available
%as we will see in the next subsection.
%We then come back to the case where $m_\pi L \simle 5$ to
%estimate the ESC numerically.

\subsection{One-pion exchange potential only}

In order to gain insight and verify our numerics we first consider a case where the ESC can be computed analytically. This is the case of a potential including {\it only} the one-pion exchange piece. Without the short range part the nuclear potential becomes rather weak in this channel and some approximations are possible.
The $^1S_0$-projected OPEP is given by
\begin{equation}
V^{\rm ope}(k,q) = -\frac{g_A^2 m_\pi^2}{8f_\pi^2 qk}
\ln\Big[\frac{m_\pi^2+(q+k)^2}{m_\pi^2+(q-k)^2}\Big],
\end{equation}
where $f_\pi=132$\,MeV and $g_A=1.26$.
Suppose the potential is weak enough to approximate 
$K\simeq \mathbb{K}\sim M V^{\rm ope}$. Then
$\Delta K\simeq \mathbb{F}\sim (V^{\rm ope})^2$ from Eq.~\eqref{mathbb_K}.
So, $\Delta K/K$ is roughly proportional to $V^{\rm ope}$.
Since $|V^{\rm ope}|$ is monotonically decreasing with $k$, 
the ESC is largest when $k\rightarrow 0$, which is our point of interest.
Below we show an analytical form of the ESC associated with the OPEP 
in low energy scatterings.

We rewrite the $\mathbb{F}$-matrix defined in Eq.~\eqref{mathbb_F} in a different form suitable for the following discussion of numerical estimates.
\begin{equation}
\mathbb{F}(p,k) = 
\sum_{|\vec{m}|\ne0,\vec{m}\in\mathbb{Z}^3}^{|\vec{m}|\le\Lambda_m}
\mathbb{F}^{|\vec{m}|}(p,k) = \sum_{n=1}^{n\le\Lambda_m^2} c_n
\mathbb{F}^{\sqrt{n}}(p,k), ~~\Lambda_m \rightarrow \infty,
\label{eq:F}
\end{equation}
where
\begin{equation}
\mathbb{F}^{|\vec{m}|}(p,k)=-M \int \frac{dq q}{2\pi^2}
\frac{\sin(|m|Lq)}{|\vec{m}|L}
\frac{V(p,q)\mathbb{K}(q,k)-V(p,k)\mathbb{K}(k,k)}{q^2-k^2},
\label{eq:partial_F}
\end{equation}
and $c_n$ is the number of distinct 3D integer vectors that share 
a common norm $\sqrt{n}$.
The cubic multiplicity $c_n$ is provided in Table~\ref{table:multiplicity} for 
$n=1, \cdots, 10$.
% % % % % % % %
\begin{table}[b]
\caption{First ten numbers of the cubic multiplicity $c_n$, introduced in 
Eq.~\eqref{eq:F}.}
\label{table:multiplicity}
\begin{tabular}{|c|cccccccccc|}
\hline
$n$   & 1 &  2 & 3 & 4 &  5 &  6 & 7 &  8 &  9 & 10 \\
$c_n$ & 6 & 12 & 8 & 6 & 24 & 24 & 0 & 12 & 30 & 24 \\
\hline
\end{tabular}
\end{table}
% % % % % % % %

\begin{figure}[t]
\begin{tabular}{cc}
\ig[width=0.48\tw]{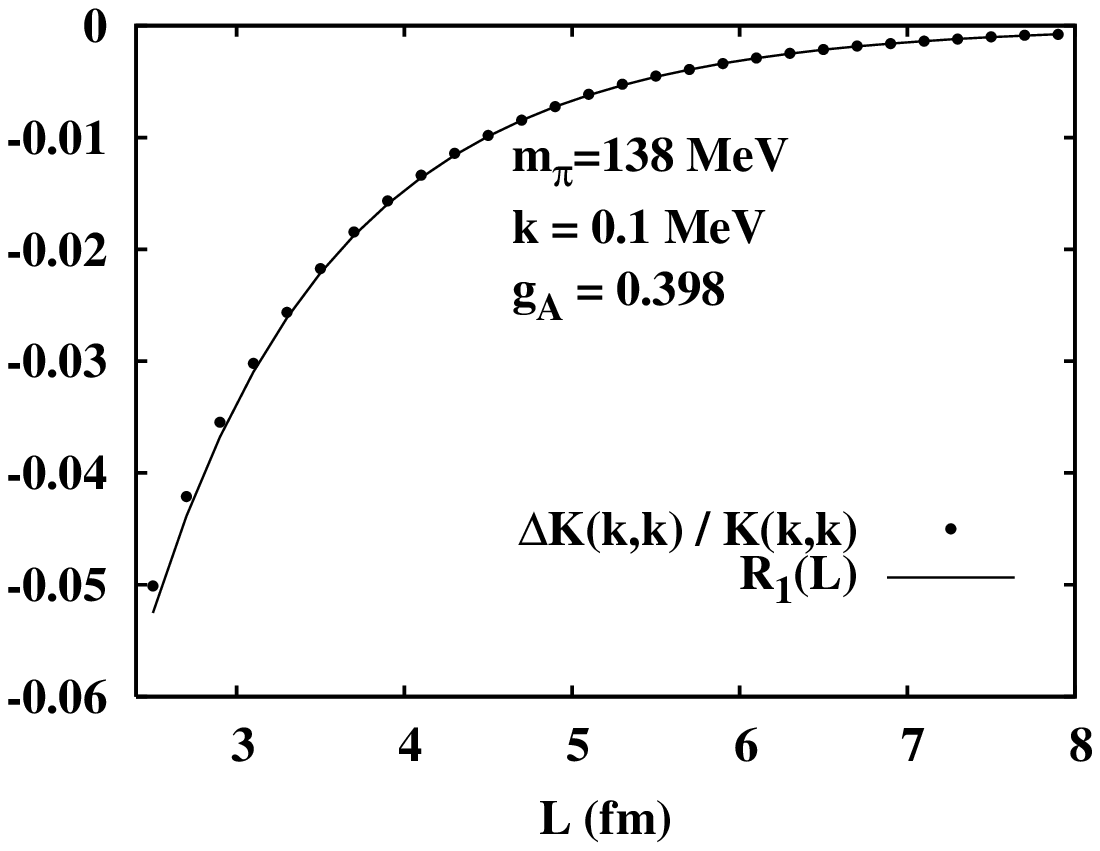} &
\ig[width=0.48\tw]{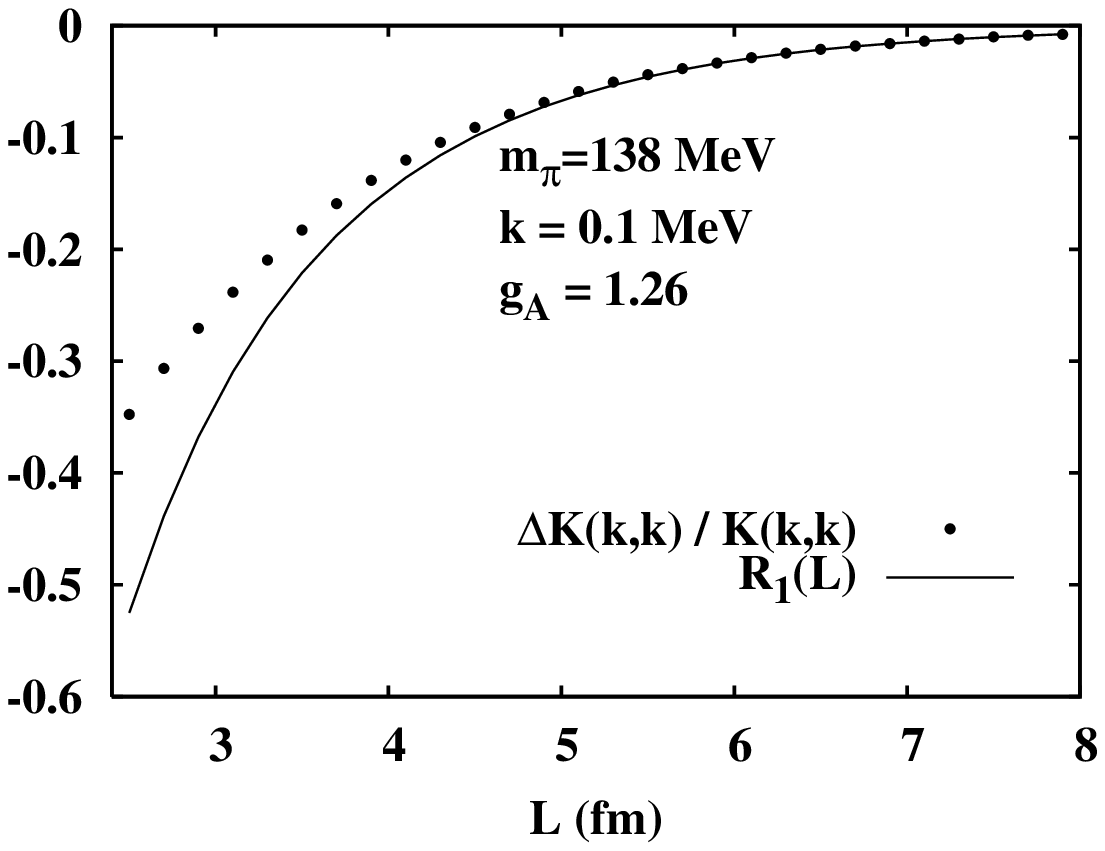} \\
(A) & (B)
\end{tabular}
\caption{Plots of $\Delta K(k,k)/K(k,k)$ as a 
function of $L$ using the OPEP (no contact potentials).
Two values of $g_A$ constant are used:
(A) $g_A=1.26/\sqrt{10}$ and (B) $g_A=1.26$.
In the case  
(A) $|MV(k,k)-K(k,k)|/|K(k,k)|\simeq 0.023$, 
and in the case  
(B) $|MV(k,k)-K(k,k)|/|K(k,k)|\simeq 0.24$.
In the calculation of $\Delta K$ only the $\mathbb{F}^1$ is used 
as it gives the largest ESC.
The cutoff in the potential is set to 500\,MeV, 
the pion mass is set to 138.0\,MeV,
and external momentum is set $k=0.1$\,MeV.
The function $R_1(L)$ computed using the same parameters as the data sets 
is drawn in each figure.}
\label{fig:opep}
\end{figure}

The OPEP is expanded in terms of $k$ as
\begin{equation}
V^{\rm ope}(k,q)
=-\frac{g_A^2 m_\pi^2}{2 f_\pi^2} \frac{1}{m_\pi^2+q^2} + O(k^2).
\label{opep_expansion}
\end{equation}
%The potential is dominated for $q\simeq 1/m_\pi$ so the potential regulator 
%plays little role.
From Eq.~\eqref{mathbb_F}, the ESC term $\mathbb{F}^{\sqrt{n}}$ corresponding to the lowest order of the OPEP in Eq.~\eqref{opep_expansion} is
\begin{align}
\mathbb{F}^{\sqrt{n}}(k,k)
\simeq& -\frac{M^2}{2\pi^2 \sqrt{n}L}
\int_0^\infty dq \frac{q\sin(\sqrt{n}Lq)}{q^2-k^2}
[(V^{{\rm ope}}(k,q))^2-(V^{{\rm ope}}(k,k))^2] \nn\\
\simeq& -\frac{g_A^4 m_\pi^4 M^2}{8\pi^2 f_\pi^4 \sqrt{n}L}
\int_0^\infty dq \frac{\sin(\sqrt{n}Lq)}{q}
\Big[\frac{1}{(m_\pi^2+q^2)^2}-\frac{1}{m_\pi^4}\Big], \nn
\end{align}
where in the first line $\mathbb{K}\simeq MV^{\rm ope}$ approximation is
used.
The integral can be performed analytically and the result is
\begin{align}
\int_0^\infty dq \frac{\sin(\sqrt{n}Lq)}{q} 
\Big[\frac{1}{(m_\pi^2+q^2)^2}-\frac{1}{m_\pi^4}\Big]
=& \frac{\pi}{2m_\pi^4}\Big[1-\frac{1}{2}e^{-\sqrt{n}m_\pi L}
(2+\sqrt{n}m_\pi L)\Big] - \frac{\pi}{2m_\pi^4} \nn\\
=&-\frac{\pi}{4m_\pi^4} e^{-\sqrt{n}m_\pi L}(2+\sqrt{n}m_\pi L)
~~\mbox{\footnotesize } \nn
\end{align}
from formula [3.735] in Ref.~\cite{Gradshteyn}.
Therefore, one obtains 
\begin{equation}
\mathbb{F}^{\sqrt{n}}(k,k)=\frac{g_A^4 m_\pi M^2}{32\pi f_\pi^4}
\frac{e^{-\sqrt{n}m_\pi L}}{\sqrt{n}m_\pi L}(2+\sqrt{n}m_\pi L) + O(k^2).
\label{eq:F1}
\end{equation}
Using only the OPEP, we numerically compute $\Delta K(k,k)/K(k,k)$ from Eqs.~\eqref{eq:KK}, \eqref{mathbb_K}, \eqref{mathbb_F}, \eqref{DK} using the ${\mathbb F}=6{\mathbb F}^1$ approximation, and plot it as a function of $L$ with fixed $k\ll m_\pi$ as shown in Fig.~\ref{fig:opep}. 
% % % % % % % %
% % % % % % % %
\begin{figure}[t]
\ig[width=0.48\tw]{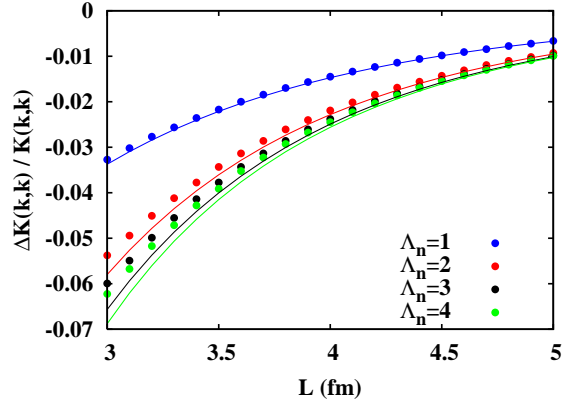}
\caption{Plots of $\Delta K(k,k)/K(k,k)$ as a 
function of $L$ using the OPEP (no contact potentials).
Here only the partial sum of the F-matrix
$\sum_{n=1}^{n\le\Lambda_n}c_n \mathbb{F}^{\sqrt{n}}$ are used with
$\Lambda_n=1,2,3,4$.
The potential is weakened by setting $g_A=1.26/\sqrt{10}$.
The cutoff in the potential is set to 500\,MeV,
the pion mass is set to 138.0\,MeV,
and the external momentum is set $k=0.1$\,MeV.
Four corresponding curves are drawn in the figure:
$R_1(L), R_2(L), R_3(L), R_4(L)$.}
\label{fig:opep_sqrt}
\end{figure}
% % % % % % % % %
In the figure we plot the function $R_n(L)$, 
which has the definition
\begin{align}
R_n(L) \equiv& \sum_{n'=1}^{n'\le n}
\frac{c_{n'} \mathbb{F}^{\sqrt{n'}}(0,0)\vert_{{\mathbb K}=MV^{\rm ope}}}{M V^{{\rm ope}}(0,0)} \nn \\
=& -\sum_{n'=1}^{n'\le n}
\frac{c_{n'} g_A^2 m_\pi M}{16\pi f_\pi^2} 
\frac{e^{-\sqrt{n'}m_\pi L}}{\sqrt{n'}m_\pi L}
(2+\sqrt{n'}m_\pi L),
\label{eq:R}
\end{align}
with $n=1$.
For a weak potential approximation, $\Delta K \simeq {\mathbb F}$
and ${\mathbb K}\simeq MV$.  Therefore the function $R_1(L)$ represents an
approximated curve for the $\Delta K(k,k)/K(k,k)$
for weak potential and low-energy scattering.
The exponential suppression $e^{-m_\pi L}$ comes from the first term
in the integer vector summation, $\sum_{\vec{m}\neq 0} \sin(|\vec{m}|Lq$),
which appears in the definition of the $\mathbb{F}$-matrix.
The second, third, and $n$-th terms
correspond to corrections proportional to 
$e^{-\sqrt{2}m_\pi L}$, $e^{-\sqrt{3}m_\pi L}$, and 
$e^{-\sqrt{n}m_\pi L}$, respectively.
Finite volume corrections $\Delta K(k,k)/K(k,k)$ using the ``partial sums'' 
of the $\mathbb{F}$-matrix, 
$\sum_{n=1}^{n\le\Lambda_n}c_n \mathbb{F}^{\sqrt{n}}$
for finite $\Lambda_n$,
are plotted in Fig.~\ref{fig:opep_sqrt} for $\Lambda_n=1,2,3,4$.
Each figure also contains a function $R_{\Lambda_n}(L)$.
As is clear in the figure, the effects of $\mathbb{F}^{\sqrt{n}}$ 
with higher $n$ are smaller than those of the lowest few partial-$\mathbb{F}$,
though the convergence of the sum of series is slow.

The unrealistic case discussed in this section shows that, as expected, the leading ESC are proportional to $e^{-m_\pi L}$. It also reinforces our confidence in the numerics involved.
%We only need to estimate the order of magnitude of the quantity 
%$\Delta K(k,k)/K(k,k)$, so in this work estimation of 
%the first term with $n=1$ is sufficient.

\subsection{Realistic potential and pion mass}

In order to estimate the ESC in the realistic case we use the potential discussed above with $C_0$ and $C_2$ being of order $4\pi/M\Lambda$ and $4\pi/M\Lambda^3$, respectively. There is some latitude in the fitting procedure used to determine the short distance constants $C_0$ and $C_2$. In order to explore the sensitivity of our results to the choice of constants we selected many sets of $(C_0, C_2)$, which reproduce physical phase shifts for $k$ up to a few hundred MeV.
We determined the value of $k$ solving the \Lu's formula (Eq.~\ref{eq:luscher}) and use this value to compute $\mathbb{F}(k,k)$ and the deviation in the phase shift
$\Delta\delta(k)$ due to the ESC, where $\Delta\delta(k)$ is defined as
\begin{equation}
k \cot [\delta(k)+\Delta\delta(k)] = -\frac{4\pi}{\mathbb{K}(k,k)}.
\end{equation}

%%%%%%%%%%%%%%%%%%%%%%%%%%%%%%%%%%%%%%%%%
\begin{figure}[t]
\ig[width=0.48\tw]{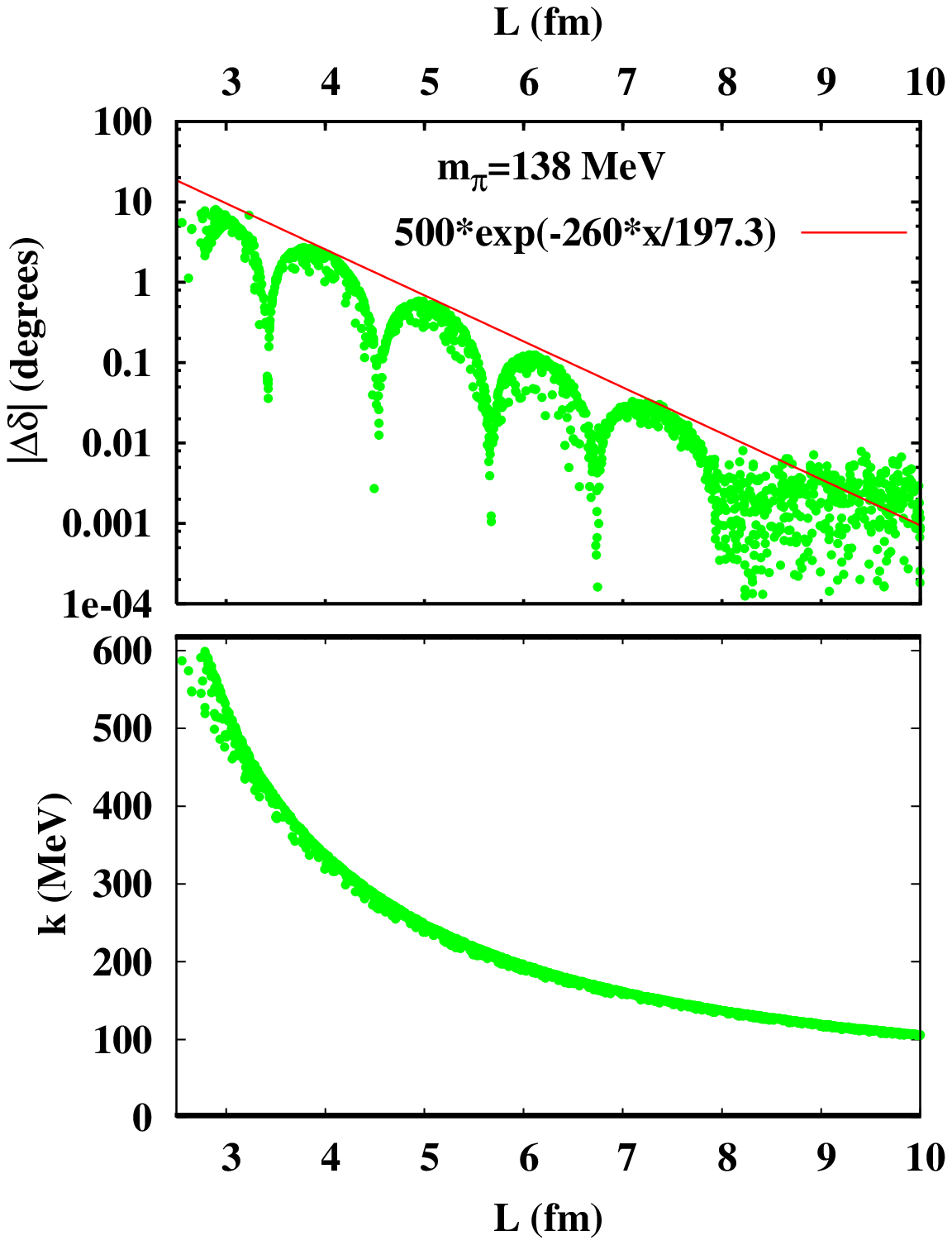}
\caption{The upper figure shows the deviation in the phase shift due to the ESC as a function of box size $L$. Realistic pion masses were used.  Contact potential parameters $C_0$ and $C_2$ are randomly selected but are constrained in such a way that they reproduce empirical phase shifts and stay within dimensional analysis range.
The ${\mathbb{F}}$-matrix is approximated to be $6{\mathbb{F}}^1$ as it gives the largest ESC (see Eq.~\eqref{eq:F}).  The ESC were calculated at the value of $k$ determined by the \Lu's formula. The lower figure shows values of the lowest (real) $k$ as a function of $L$, determined from the \Lu's formula.}
\label{fig:realistic}
\end{figure}
\begin{figure}[t]
\ig[width=0.48\tw]{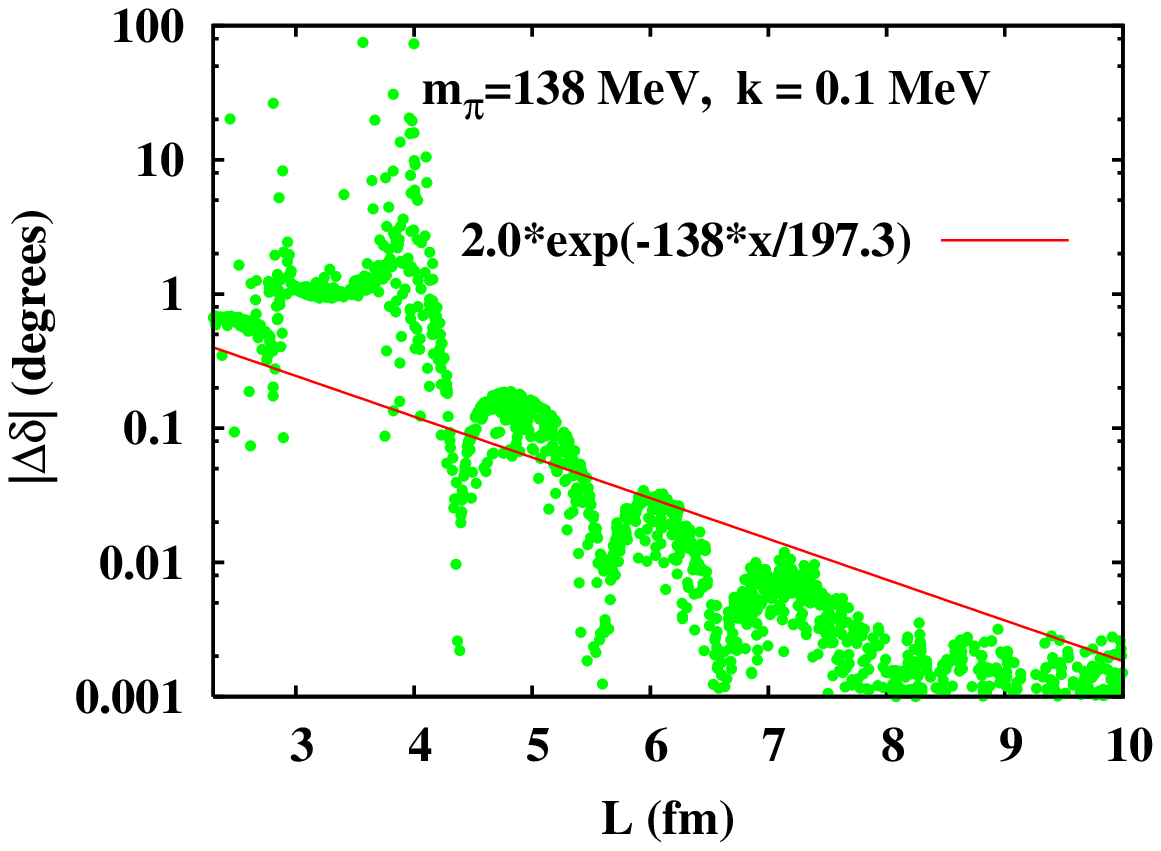}
\caption{Deviation in the phase shift due to the ESC as a function of box size $L$. Realistic pion masses were used and the ESC were calculated at fixed value of $k=0.1$\,MeV. Contact potential parameters $C_0$ and $C_2$ are randomly selected but are constrained in such a way that they reproduce empirical phase shifts and stay within dimensional analysis range. The ${\mathbb{F}}$-matrix is approximated to be $6{\mathbb{F}}^1$ as it gives the largest ESC (see Eq.~\eqref{eq:F}).}
\label{fig:realistic_kfixed}
\end{figure}

%%%%%%%%%%%%%%%%%%%%%%%%%%%%%%%%%%%%%%%%%
The result is plotted, as a function of the box size $L$ in Fig.~\ref{fig:realistic}. As expected the corrections are small and decrease with box size, approximately in an exponential fashion. The ESC shown in Fig.~\ref{fig:realistic} however do not follow naively expected exponential behavior, $\sim e^{-m_\pi L}$ (with characteristic scale being $m_\pi$),  but instead, numerical results indicate the characteristic scale being $\simeq 260$\,MeV.
Both $k$ and $m_\pi$ combine to set the scale for the exponential decay of finite volume corrections with increasing box size. To verify this statement we repeated the calculation holding $k$ fixed and small ($k=0.1$\,MeV). As Fig.~\ref{fig:realistic_kfixed} shows, the agreement between the numerical results and the expected $e^{-m_\pi L}$ behavior is roughly observed.
%%%%%%%%%%%%%%%%%%%%%%%%%%%%%%%%%%%%%%%%%
%%%%%%%%%%%%%%%%%%%%%%%%%%%%%%%%%%%%%%%%%

\subsection{Extrapolation to higher pion masses}

\begin{figure}[t]
\ig[width=0.48\tw]{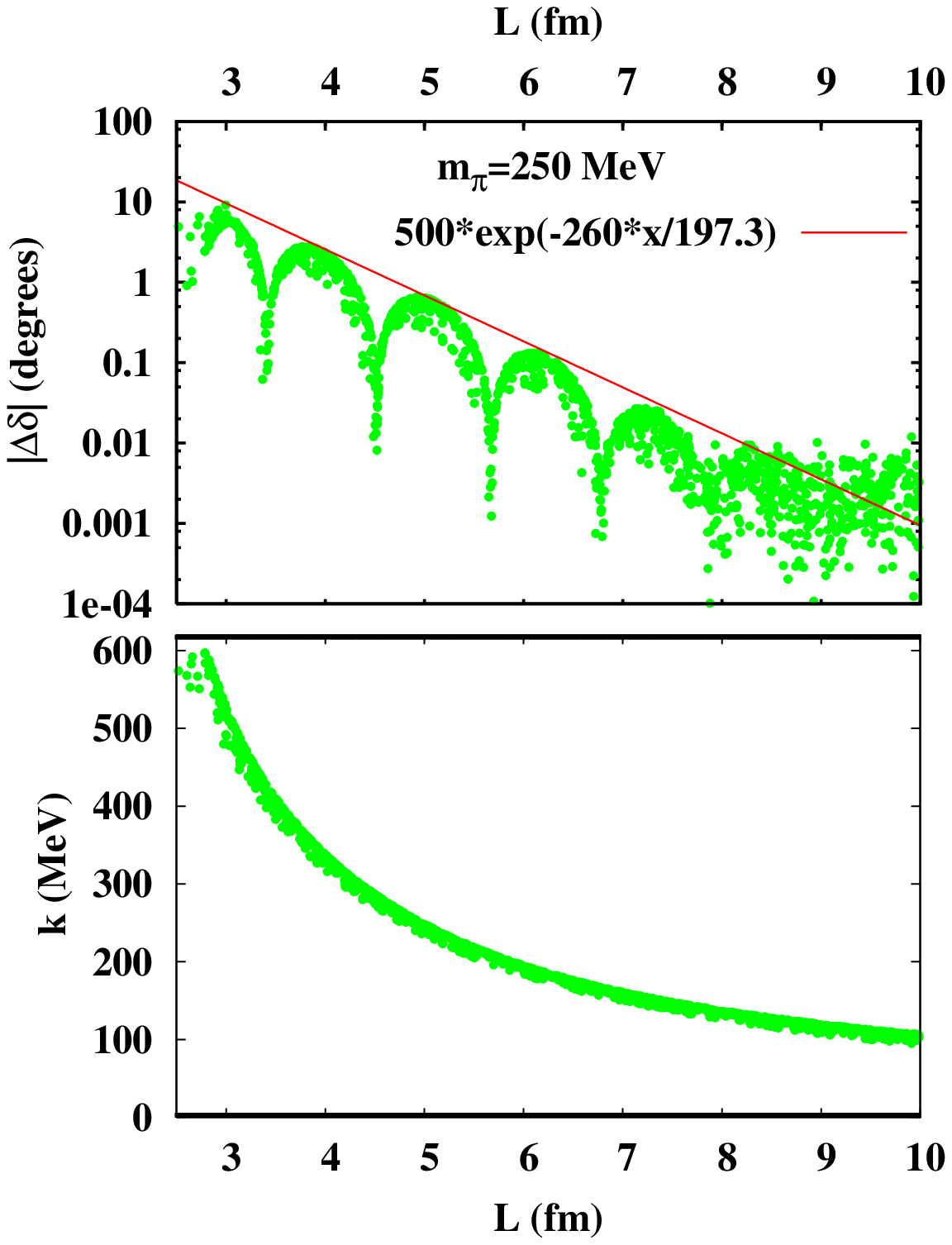}
\caption{The upper figure shows the deviation in the phase shift due to the ESC as a function of box size $L$. Pion masses were set at $250$\,MeV.  Contact potential parameters $C_0^0$, $C_0^2$ and $C_2$ are randomly selected as described in the text. The ${\mathbb{F}}$-matrix is approximated to be $6{\mathbb{F}}^1$ as it gives the largest ESC (see Eq.~\eqref{eq:F}). The ESC were calculated at the value of $k$ determined by the \Lu's formula. The lower figure shows values of the lowest (real) $k$ as a function of $L$, determined from the \Lu's formula.}
\label{fig:250}
\end{figure}

In order to estimate the ESC in current lattice calculations we need to compute them for pion masses used in these calculations. The nuclear potential is sensitive to the value of the pion mass so, as mentioned before, it is difficult to predict what the nuclear potential at higher pion masses is. One might expect however that, as the pion mass grows and the range of the nuclear force decreases, the ESC would decrease. In fact, if the value of $k$ coming from the solution of the \Lu's formula is used, it sets the scale for the exponential decay of the ESC with $L$ and there is little difference between the ESC computed with $m_\pi=138$\,MeV and $m_\pi=250$\,MeV. This is exemplified in the Fig.~\ref{fig:250}. 
%%%%%%%%%%%%%%%%%%%%%%%%%%%%%%%%%%%%%%%%%
%%%%%%%%%%%%%%%%%%%%%%%%%%%%%%%%%%%%%%%%%
The extrapolation is done by first choosing $C_0$ and $C_2$, which reproduce empirical phase shifts, and then choosing $C_0^0$ and $C_0^2$ constrained in the range $|C_0^2| m_\pi^2 < (1/15) |C_0^0|$ so that these coefficients are within their naive dimensional analysis ranges.  Choices of $(C_0, C_2)$ and $(C_0^0, C_0^2)$ are randomly made under the constraints.  In this calculation, the values of $M, f, g_A$ are held fixed at physical values.

%\subsection{Nucleon mass in finite volume}
%
%In addition to the corrections to the scattering matrix, there is 
%also a correction given by renormalization of nucleon mass in finite 
%volume.  Nucleon mass is shifted in finite volume due to loop
%effects in the chiral expansion.  This mass gap $\delta M=M_L-M_\infty$ affects
%the inhomogeneous term appearing in the defining function of $\mathbb{K}$.
%For nonzero $\delta M$, Eq.~\eqref{mathbb_K} becomes
%\begin{equation}
%\mathbb{K}(p,k) = M_\infty V(p,k) - \pr \frac{d^3q}{(2\pi)^3} 
%\frac{M_\infty V(p,q)\mathbb{K}(q,k)}{q^2-k^2} + 
%[\delta M V(p,k) + \mathbb{F}(p,k)].
%\end{equation}
%Mass shift appearing in the integral is neglected as this effect is of higher
%order.  The asymptotic form of $\delta M$ is given by~\cite{Beane:2004tw}
%\begin{equation}
%\delta M=\frac{9 g_A^2 m_\pi^2}{8\pi f_\pi^2} \frac{e^{-m_\pi L}}{L},
%\end{equation}
%where we only use the term proportional to $g_A^2$ and neglect the term
%proportional to $g_{\Delta N}^2$.
%So, the ESC originated from the mass shift is 
%\begin{equation}
%|\delta M V(0,0)|=\frac{9g_A^4 m_\pi^2}{16\pi f_\pi^4} \frac{e^{-m_\pi L}}{L},
%\end{equation}
%while the leading ESC from $\mathbb{F}$ is given by Eq.~\eqref{eq:F1} for
%$n=1$.  Apparently, for low-energy scatterings $\mathbb{F}^1>|\delta M V|$ 
%roughly by a factor of $\frac{M^2}{m_\pi^2}(1+m_\pi L/2)$, 
%which is usually substantially large.
%Therefore, it is safe to disregard the nucleon mass gap in the estimates of
%ESC in finite volume scatterings.

\section{Conclusion}

We have derived the relation between energy levels and phase shifts keeping terms that are exponentially suppressed in the large volume limit but that can be significant for the lattice sizes currently in use. These terms indeed appear suppressed by a factor of $e^{-m_\pi L}$ as long as they are computed at energy much smaller than the pion mass. In actual lattice calculations, the value of the energy levels measured are a function of the volume used and are usually in the few hundreds of MeV range. That changes the suppression factor: it is still exponential but with a different slope than the naive estimate $\sim e^{-m_\pi L}$. We also show that the effect of pions coming from one nucleon, ``going around the lattice'' and interacting with the other nucleon vanishes in linear order in the potential if the energy shift is expressed in terms of the (infinite volume) phase shifts.

We performed numerical estimates using an effective field theory inspired potential. Other sources of exponential corrections (shifts in the nucleon mass, pion decay constant, changes in the one-loop potential) are argued to be small on the basis of the chiral expansion. Our numerical estimates indicate that, in a calculation with realistic pion masses,  a lattice size of about $(5\,\mbox{fm})^3$ is necessary (and sufficient) for these corrections to stay within $1^\circ$. Contrary to naive expectations an increase of the pion mass does not substantially reduces the minimum box size necessary for a given precision level.

\begin{acknowledgements}
The authors would like to thank Thomas Cohen  and Markus Luty for discussions on this subject.
This work is supported by the Director, Office of Energy Research, Office of High Energy and Nuclear Physics, Divisions of Nuclear Physics, of the U.S. Department of Energy under Contract No. DE-F603-ER-40762 and DE-AC02-05CH11231.
\end{acknowledgements}

%\section{Appendix}
%I'll keep here for now the extrapolations in the potential parameters used in the numerics.
%
%\beq
%f(m_\pi) = f\left[ 1+ \frac{m_\pi^2}{8\pi^2 f^2 \log\left(\frac{\Lambda_4^2}{m_\pi^2}\right)}\right],
%\eeq with $f=124.8$ MeV and $\Lambda_4=942.8$ Mev.
%
%\beq
%M(m_\pi) = M-4 c_1 m_\pi^2-\frac{3g^2 m_\pi^3}{32\pi f^2},
%\eeq with$M=919.3$ MeV, $c_1=-.00060$ and
%
%\beq
%g_A(m_\pi) = g-\frac{g g_{N\Delta}^2}{4\pi^2f^2} J
%-\frac{25 g_{\Delta\Delta} g_{N\Delta}^2 }{324\pi^2 f^2} J
%+\frac{2gg_{N\Delta}^2}{9\pi^2f^2} K
%-\frac{g}{8\pi^2 f^2} L
%-\frac{g^3}{4\pi^2 f^2}L + m_\pi^2 C,
%\eeq with
%\bea 
%J &=& (m_\pi^2-2 \Delta^2)\log(m_\pi^2/\mu^2)+2\Delta\sqrt{\Delta^2-m_\pi^2}
%\log\left( \frac{\Delta-\sqrt{\Delta^2-m_\pi^2}}{\Delta+\sqrt{\Delta^2-m_\pi^2}}\right)\nn\\
%K &=& (m_\pi^2-2/3 \Delta^2)\log(m_\pi^2/\mu^2)
%+\frac{2}{3}\Delta\sqrt{\Delta^2-m_\pi^2}
%\log\left( \frac{\Delta-\sqrt{\Delta^2-m_\pi^2}}{\Delta+\sqrt{\Delta^2-m_\pi^2}}\right)\nn\\
%&+&\frac{2}{3}\frac{m_\pi^3}{\Delta}(\pi m_\pi-\sqrt{\Delta^2-m_\pi^2}
%\log\left( \frac{\Delta-\sqrt{\Delta^2-m_\pi^2}}{\Delta+\sqrt{\Delta^2-m_\pi^2}}\right)\nn\\
%L &=& m_\pi^2 \log(m_\pi^2/\mu^2)
%\eea and $\Delta=271$ MeV, $g=0.932$, $g_{N\Delta}=-1.43$, $g_{\Delta\Delta}=-3.72$, $C=-6.43$ and $\mu=1000$ MeV. This complicated formula for $g_A$ is numerically the same, within the errors,  as
%\beq
%g_A(m_\pi) \approx 1.29-0.00017 \frac{m_\pi}{MeV}.
%\eeq Excepto for the last one, all these formulae are one-loop chiral pert. theory results fit to lattice data.

\newpage

\end{document}